\documentclass{elsart}

\usepackage{graphicx}

\usepackage{amssymb}

\begin{document}

\begin{frontmatter}



\title{
Localization of Electronic States in Chain Models Based on Real DNA Sequence
}



 \author[yamada]{Hiroaki Yamada}
 \address{
Aoyama 5-7-14-205,
Niigata 950-2002,
Japan
}

\thanks[yamada]{Corresponding author, 
electronic mail address: hyamada@uranus.dti.ne.jp, 
Former address: Faculty of Engineering, 
 Niigata University, Ikarashi 2-Nocho 8050, Niigata 950-2181, Japan.
}

\begin{abstract}
We investigate 
the localization property of an electron 
in the disordered two- and three-chain systems (ladder model) 
with long-range correlation
as a simple model for electronic property in a double strand of DNA.
The chains are constructed by repetition 
of the sugar-phosphate sites, and the inter-chain hopping 
at the sugar sites come from nucleotide pairs, i.e., $A-T$ or $G-C$ pairs.   
It has been found that some DNA sequences have long-range correlation.
In this paper we investigate the localization properties of the electronic states in 
 some actual DNA sequences such as bacteriophages of escherichia coli,  
 human chromosome 22 and histone protein. 
    We will present some numerical results for 
the Lyapunov exponent (inverse localization length) of the wave function
in the cases in comparison to the results for artificial sequence
generated by  an asymmetric  modified Bernoulli map. 
It is shown that the correlation and asymmetry of the sequence affect on 
the localization in both the artificial and the real DNA sequences.
\end{abstract}


\end{frontmatter}

\section{Introduction}
The recent development of the nanoscale fabrication let us expect
the utilization of the DNA wire as a molecular device \cite{lewis03,porath04} 
and the realization of DNA computing \cite{paun98}.
Actually, the modern development enables us
to measure the direct DNA transport phenomena \cite{porath00,tran00}.
   Recently, Porth {\it et al.} measured the  
nonequilibrium current-voltage ($I-V$) characteristics in the
poly(G)-poly(C) DNA molecule attached to platinum lead at room temperature 
\cite{porath00}. 
 Cuniberti {\it et al.} explained the semiconducting behavior by considering
the base-pair stack coupled to the sugar-phosphate (SP) backbone pair
\cite{cuniberti02}.
 Iguchi also derived the semiconductivity and the band gap by using 
the ladder chain model of the double strand of DNA \cite{iguchi97}.
 In above models, the existence of the SP backbone chains  
play an important role in the band structure due to 
the gap opening
 by the hybridization of the energy levels.
Furthermore, recent ab-initio calculations in short segments show that the backbone chain of DNA
might play an imprtant role for the entire electronic spectrum of the system
 \cite{kurita00,felice01}. 

 On the other hand, 
 Tran {\it et al.} measured the conductivity along the lambda phage DNA ($\lambda-$DNA)
 double helix at microwave frequencies using the lyophilized DNA in and also 
without a buffer \cite{tran00}. 
The conductivity is strongly temperature dependent around room temperature with 
a crossover to a weakly temperature dependent conductivity at low temperatures.
 Yu and Song showed that the observed temperature dependent
conductivity in the DNA can be consistently
modeled, without invoking the additional ionic conduction
mechanisms, by considering that electrons may use
the variable range hopping for conduction and that electron
localization is enhanced by strong thermal structural
fluctuations in DNA \cite{yu01}.
  Then the DNA double helix is viewed as a one-dimensional Anderson system.
Carpena {\it et al.} and Roche used some real DNA sequence as the on-site energy 
in the tight-binding one-dimensional system 
to investigate the localizaton property of the wavefunctions  
\cite{carpena02,roche03}.

The transport property though DNA are still controversial mainly
due to the tremendous difficulties in the setting up the proper experimental environment
and the DNA molecule itself.
 Although many theoretical explanations for the charg transport phenomena have been suggested
on the basis of the standard solid-state-physics approach such as 
polarons, solitons, hole hopping model on guanine sites 
\cite{iguchi03,iguchi03a,hennig02,ladik99,shen02,bruinsma00,kats02,starikov02,damle02,troisi02,rodriguez03,berlin02,lebard03,lewis03},
the situation has been  still far from unifying the theoretical scheme.

 Moreover, as one of the realistic situation, it has been found that the base (nucleotide)
sequence of the various genes 
has long-range correlation 
characterized by the power spectrum 
$S(f)\sim f^{-\alpha}$ ($0.1<\alpha <0.8$) in the low frequency limit ($f<<1$)
\cite{voss92,buldyrev95,bishop97,holste01,holste03,isohata03,grosse02}. 
As observed in the power spectrum, the mutual information analysis 
and the Zipf analysis of the DNA base
sequence such as the human chromosome 22 (HCh-22) , the long-range structural correlation exists in the 
total sequence as well as the short-range periodicity \cite{holste01,holste03,isohata03,grosse02}.   
 Eukaryote's DNA sequence has apparently periodic repetition in terms of the gene duplication.
The correlation length in the base sequence of genes changes from the early eukayote
to the late eukaryote by the evolutionary process. 
It seems that the long-range correlations tends to manifest in power spectra of the total sequences rather 
that in those of the exon part and the intron part separately \cite{isohata03}.

The localization property of the single-chain disordered system with long-range
correlation has been extensively studied \cite{yamada91,moura98,yamada03}.
Accordingly, to compare the localization nature of the electronic states in the real DNA sequence
with that in the disordered sequence with long-range correlation is very interesting.
   In the present paper, we numerically give localization nature of the electronic states 
in some real DNA sequences
such as bacteriopages of escherichia coli (E. coli),  HCh-22 and histon H1. 
We also investigate the correlation effect on 
the localization property of the one-electronic states 
in the disordered  chain models  
with a long-range structural correlation.
We present some numerical results for 
the Lyapunov exponents of the wave function.
 In particular, it is found that the correlation of the sequence enhances the 
localization length and asymmetry of the distribution of the elements in the sequence affect
on the localization.

Note that the real values for the biological molecule 
such as the ionization energy \cite{ladik88}, the electronegativity and so on, 
are not used in the numerical calculation in the present paper. We used the simpler values as 
the parameters in order to show some basic localization 
and delocalization properties of the electronic states in the ladder models.
We would like to mainly focus on (1)suggesting the model and (2)giving
the preliminary numerical results of the electronic  localization 
in the model with real DNA sequences.

  Outline of the present paper is as follows.
In the next section we introduce the simple model for DNA in order to investigate the electronic states.
In the Sec. 3 we give the mapping rules of the real DNA sequences and an asymmetric modified
Bernoulli map in order to generate artificially the  correlated sequences.
The numerical results for the Lyapunov exponent and the localization length 
in the systems are given in Sect. 4. 
The last section contains summary and discussion.

\section{Model}
We simplify and model the double strand of DNA by some assumptions.
DNA double helix structure is constructed by the coupled two single strand of 
DNA.
 First, we ignore the twist of DNA as well as the complicated topology. 
In addition to the simplification, we consider only the $\pi-$electrons in the backbones and 
the base-pairs of the system. We also ignore the interaction 
between the electrons and restrict ourselves to the zero-temperature property.

Following the basic assumptions, 
consider the one-electron system described by  
the tightly binding model consisting of the 
two- or three- chains.
The ladder model  was first introduced by Iguchi \cite{iguchi97}
as a model for considering the electronic properties of
a double strand of DNA.
   The Schr\"{o}dinger equation is given as,
\begin{eqnarray*}  
A_{n+1,n} \phi^{A}_{n+1}+ A_{n,n-1} \phi^{A}_{n-1} + A_{n,n} \phi^{A}_{n} 
+ V_{n} \phi^{B}_{n}  = E\phi^{A}_{n}, \\ 
B_{n+1,n} \phi^{B}_{n+1}+ B_{n,n-1} \phi^{B}_{n-1} + B_{n,n} \phi^{B}_{n} 
+ V_{n} \phi^{A}_{n}  = E\phi^{B}_{n},  
\end{eqnarray*}  
\noindent
where
$A_{n+1,n}$ ($B_{n+1,n}$) means the hopping integral 
between the $n$th and $(n+1)$th sites and $A_{n,n}$ ($B_{n,n}$) 
the on-site energy at site $n$ in chain $A$
($B$), and $V_{n}$ is the hopping integral 
from chain $A (B)$ to chain $B (A)$ at site $n$,
respectively.

\begin{figure}[!h] 
\begin{center} 
\includegraphics[scale=.65]{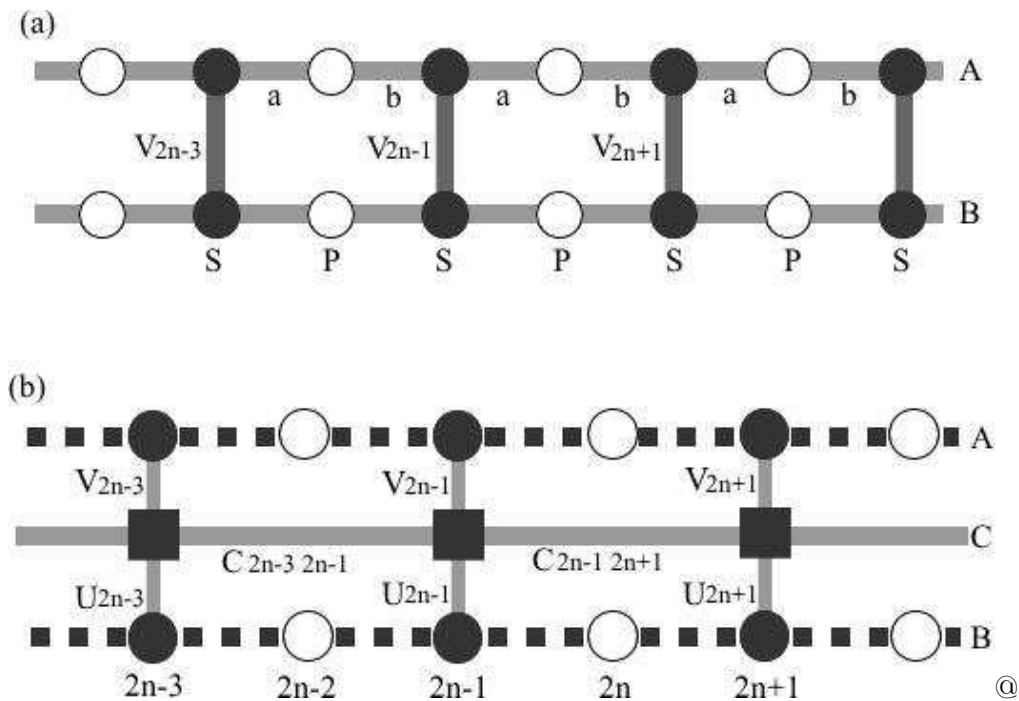}　
\caption{Models of the double strand of DNA.
(a) The two-chain model,  (b) the three-chain model, where $S$ and $P$ represent
sugar and phosphate sites, respectively. 
}
\end{center} 
\end{figure}

Furthermore it can be rewritten in the matrix form,
\begin{eqnarray*}  
 \left(
\begin{array}{c}
\phi^{A}_{n+1} \\
\phi^{B}_{n+1} \\ 
\phi^{A}_{n} \\ 
\phi^{B}_{n} 
\end{array}
\right)
=
\left(
\begin{array}{cccc}
\frac{E-A_{nn}}{A_{n+1n}} & -\frac{V_n}{B_{n+1n}} & -\frac{A_{n-1n}}{A_{n+1n}} & 0 \\
-\frac{V_n}{B_{n+1n}}  & \frac{E-B_{nn}}{B_{n+1n}} & 0& -\frac{B_{n-1n}}{B_{n+1n}}  \\
 1 & 0 & 0 & 0 \\
0 & 1 & 0 & 0 
\end{array}
\right)
\left(
\begin{array}{c}
\phi^{A}_{n} \\
\phi^{B}_{n} \\ 
\phi^{A}_{n-1} \\ 
\phi^{B}_{n-1} 
\end{array}
\right) 
\equiv T_{d=2}(n)
\left(
\begin{array}{c}
\phi^{A}_{n} \\
\phi^{B}_{n} \\ 
\phi^{A}_{n-1} \\ 
\phi^{B}_{n-1} 
\end{array}
\right).
\end{eqnarray*}  
\noindent
  We would like to investigate the asymptotic 
behavior ($N \to \infty$) of the products of the matrices 
$M_d(n)=\Pi_k^n T_{d=2}(k)$ \cite{crisanti93}. 
According to the parameter sets given by Iguchi \cite{iguchi97},
we set $A_{n+1,n} = B_{n+1,n}=a(=b)$ at odd (even) site $n$, 
respectively and $V_n=0$ at even sites (phosphate sites)
 for simplicity. 
The chain $A$ and $B$ are constructed by the repetition 
of the sugar-phosphate sites, and the inter-chain hopping $V_n$
at the sugar sites come from the nucleotide base-pairs, i.e., $A-T$ or $G-C$ pairs.   
(See Fig.1(a).)

 The reduction to single-chain system ($d=1$) and the extension to three-chain system ($d=3$)
 are easy \cite{yamada04dna}. 
In particular, when we allow hop of the electron between the nearest  neighbor 
nucleotide base-pairs by the overlap integral as well as hop between the backbone sites,
the two-chain model can be easily extended to the three-chain one. 
The three-chaneels syatem can be described by following
Schr\"{o}dinger equation. 
\begin{eqnarray*}  
A_{n+1,n} \phi^{A}_{n+1}+ A_{n,n-1} \phi^{A}_{n-1} + A_{n,n} \phi^{A}_{n} 
+ V_{n} \phi^{C}_{n}  = E\phi^{A}_{n}, \\ 
C_{n+2,n} \phi^{C}_{n+2}+ C_{n,n-2} \phi^{C}_{n-2} + C_{n,n} \phi^{C}_{n} 
+ V_{n} \phi^{A}_{n} + U_{n} \phi^{B}_{n}   = E\phi^{C}_{n}, \\ 
B_{n+1,n} \phi^{B}_{n+1}+ B_{n,n-1} \phi^{B}_{n-1} + B_{n,n} \phi^{B}_{n} 
+ U_{n} \phi^{C}_{n}  = E\phi^{B}_{n},  
\end{eqnarray*}  
for the odd site $n$. On the other hands, 
\begin{eqnarray*}  
A_{n+1,n} \phi^{A}_{n+1}+ A_{n,n-1} \phi^{A}_{n-1} + A_{n,n} \phi^{A}_{n} 
= E\phi^{A}_{n}, \\ 
B_{n+1,n} \phi^{B}_{n+1}+ B_{n,n-1} \phi^{B}_{n-1} + B_{n,n} \phi^{B}_{n} 
= E\phi^{B}_{n},  
\end{eqnarray*}  
for the even sites $n$. 
 The geometry and setting in the three-chain model is given in Fig.1(b).

\section{Correlated sequences}
As the correlated binary sequence $\{V_n\}$ 
of the hopping integrals, we use some real DNA sequence such as the bacteriophages of 
E.coli  (phage-$\lambda$, phage-186), 
HCh-22 and histone H1.
We can get the real DNA sequence from the gene data bases \cite{dnadata}. 

When we convert the nucleotide sequences $\{ S_n \}$ to a numerical data  $\{ V_{n} \}$, some rules
are used as seen in DNA walk analyses \cite{buldyrev95}: (I)Prine-pyrimidin rule. If $S_n$ is a purine (A or G) then $V_n=W_{AG}$,
if $S_n$ is a pyrimidin (C or T) then $V_n=W_{CT}$, where $W_{AG}$ and $W_{CT}$ denote moderate
numerical values for calculation. (II)Hydrogen bond energy rule.  $V_n=W_{GC}$ for strongly bonded pairs (G-C), 
$V_n=W_{AT}$ for weakly bonded pairs (A-T), where $W_{GC}$ and $W_{AT}$ denote moderate
numerical values for calculation. (III)Hybrid rule.  $V_n=W_{AC}$ for A or C, 
$V_n=W_{GT}$ for G or T, where $W_{AC}$ and $W_{GT}$ denote moderate
numerical values for calculation.  Apparently the hydrogen bond energy rule is relevant in order to investigate 
the electronic localization in the sequence of DNA double helix. 
We apply the hydrogen bond energy rule to the real DNA sequences in the present paper.
  
Moreover, we compare the characteristics of the localization property with the result in an artificial
sequence generated by 
 following asymmetric modified Bernoulli map 
\cite{aizawa84}.

\begin{eqnarray}
 X_{n+1}= 
& \biggl( 
   \begin{array}{cc}
   X_{n} + 2^{B_0-1}X_{n}^{B_0}  & (X_{n} \in I_0)  \\
   X_{n} - 2^{B_1-1}(1-X_{n})^{B_1} & (X_{n} \in I_1),    
\end{array}
\label{eq:map}
\end{eqnarray}
\noindent
where $I_0=[0,1/2),I_1=[1/2,1)$.  
$B_0$ and $B_1$ are the bifurcation parameters which control the correlation
of the sequence, and we set $1<B_0\leq B_1<2$ for simplicity. 
The asymmetry of the map ($B_0 \neq B_1$) corresponds to the asymmetric property of
the distribution of the real sequence of the double helix DNA 
that the number of the A-T pairs does not equal one of G-C pairs 
which is different from the random binary sequence with equal weight.  
We introduce an indicator $R_{GC}$ for the rate of the G-C pair in the sequences as, 
$R_{GC}=(N_G+N_C)/(N_G+N_C+N_A+N_T)$, where $N_G, N_C, N_A$ and $N_T$ denote the number of each symbol
G,C,A and T in the sequence, respectively. 

In the ladder model we also use the symbolized sequences $\{ V_{n} \}$ and/or $\{ U_{n} \}$ 
 by the following rule as the interchain hopping integral at odd sites $n$:
\begin{eqnarray}
V_{n} =  
\left\{ \begin{array}{ll}
W_{AT}  & (X_{n} \in I_0) \\
W_{GC} & (X_{n} \in I_1). \\
\end{array} \right.
\end{eqnarray} 
\noindent
In the numerical calculation, $W_{GC}$ is set at a half of $W_{AT}$  for simplicity.
($W_{GC}=W_{AT}/2$.)
Then the artificial binary sequence can be roughly regarded as the base-pair
sequence as observed in the $\lambda-$DNA or the HCh-22. 
The correlation function $C_o(n)(\equiv <V_{n_0}V_{n_0 + n}>)$ ($n_0=1$, $n$ is even ) 
decays by the inverse power-law depending on the value $B$ as 
$C_o(n)\sim n^{-\frac{2-B_1}{B_1-1}}$ for large $n$ ($3/2<B_i<2$). 
The power spectrum becomes 
$S(f)\sim f^{-\frac{2B_1-3}{B_1-1}}$ for small $f$. 
We focus on the Gaussian and non-Gaussian stationary region ($1<B_i <2$) 
that correspond to some real DNA base-pair sequence with 
$S(f) \sim f^{-\alpha} (0.2<\alpha<1)$.
There are various ways to generate the correlated sequences as seen in study in 
one-dimensional disordered system with long-range correlation. 
We must pay attention to the statistical properties.
For example, when we use the correlated random walk with Hurst index to generate 
the correlated sequence, the sequence must be rescaled by the variance of 
the fluctuation because the fluctuation diverges with the length of the sequence. 
However, in the stationary sequence generated by the 
modified Bernoulli map the fluctuation does not diverge because it takes only
alternative values at the each sites $n$.

\section{Numerical Result} 
 
We give the numerical result of 
the energy dependence of the Lyapunov exponents.
The definition is given by,
\begin{eqnarray}  
\gamma_i = \lim_{n \to \infty} \frac{1}{2n} \log \sigma_i(M_d(n)^\dagger M_d(n)),    
\end{eqnarray}  
\noindent
where $\sigma_i(...) $ denotes the $i$th eigenvalue of the matrix $M_d(n)^\dagger M_d(n)$  \cite{yamada01}. 
 As the transfer matrix $T_d(n)$ is symplectic, the eigenvalues of the
$M_d(n)^\dagger M_d(n)$ have the reciprocal symmetry around the unity as
$e^{\gamma_1},...,e^{\gamma_d},e^{-\gamma_d},...,e^{-\gamma_1}$, 
where $\gamma_1 \geq \gamma_2\geq \dots \gamma_d \geq  0$.

Furthermore, it is found that 
for the thermodynamic limit ($n \to \infty$)
the largest channel-dependent localization length $\xi_d(=1/\gamma_d)$ 
determines the exponential decay of the Landauer conductance of the system between metallic electrodes 
as $g(n)=\sum_i (cosh(2\gamma_i n)-1)^{-1} \to \exp(-2\gamma_d n)$ in units of $2e^2/h$ at 
zero temperature and serves as the localization
length of the total system of the coupled chains \cite{crisanti93}.
Recently, the electron transport for the molecular wire between two metallic electrorodes
has been also investigated by several techniques 
\cite{tikhonov02a}.

\begin{figure}[!h] 
\begin{center} 
\includegraphics[scale=.8]{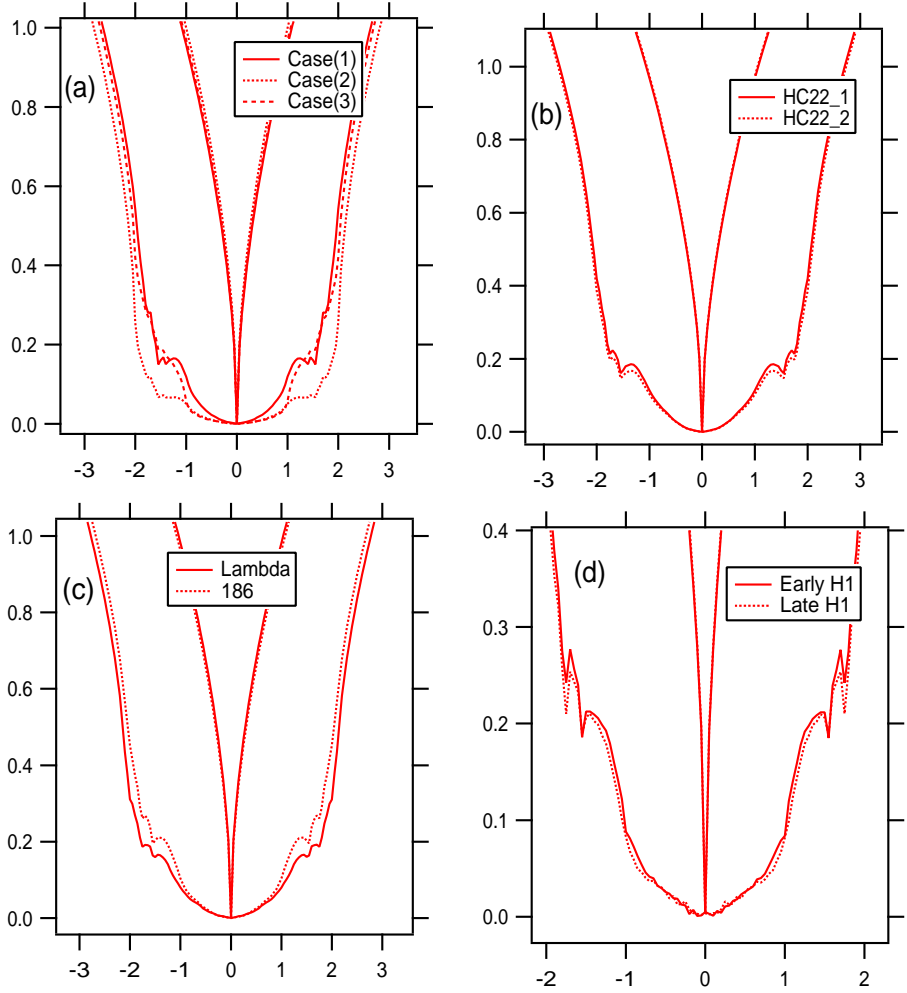}　
 \caption{Lyapunov exponents ($\gamma_1, \gamma_2$) as a function of energy in 
the ladder model. (a) modeified Bernoulli model, (b) human chromosome 22, 
(c) bacteriophages of E.coli (phage-$\lambda$, phage-186), 
 (d)early histone H1 and late histone H1.   
The on-site energy is set at $A_{nn}=B_{nn}=0, a=-1.0, b=-0.5$.
The size of the sequence is $N=10^5$ for (a), 
$N=10^5$ for (b), 
$N=48510$ for the phage-$\lambda$ in (c),
$N=30624$ for the phage-186 in (c),
$N=787$ for the early histone H1 in (d), and
$N=1182$ for the late histone H1 in (d). 
}
\end{center} 
\end{figure}

We consider the correlation effect on the localization
property of the disordered case by using some real DNA sequences and the modified Bernoulli model.
Then we used a sample with the system size $N=10^5$ for the numerical
calculation in the modified Bernoulli map.
Note that perfect periodicity exists in the deterministic (even) sites in our models. (See Fig.1.) 
We introduce another long-range correlation 
due to the base-pair sequence on the odd sites $V_{2n-1}$.

It is generally known that the correlation of the sequence enhances the delocalization in 
the electronic states. In the asymmetric modified Bernoulli system characterized by the
two-parameters, $B_0,B_1$, the correlation decay depends on $B_1$ in large sequences, because we set
$B_0\leq B_1$.    
Figures 2(a) shows the energy dependence of the Lyapunov exponents 
($\gamma_1$ and $\gamma_2$)  
for some cases in asymmetric modified Bernoulli system.
They are named as follows: 
case (1)$B_0=1.0, B_1=1.0$, case (2)$B_0=1.0, B_1=1.9$ and case (3)$B_0=1.7, B_1=1.9$. 
Apparently the case (1) is more localized than cases (2) and (3) in vicinity
 of the band center $|E|<1$.  The comparison between the case (2) and case (3)
shows the effect of the asymmetry of the map on the localization. 
$R_{GC} \sim 0.2$ for the case(2), $R_{GC} \sim 0.47$ for the case(3).
In energy regime $|E|>1$, the Lyapunov exponent $\gamma_2$ in the case (2) is smaller than
one in the case (3) in spite of the same correlation strength $B_1$.    
As a result, it is found that in the DNA ladder model the correlation and asymmetry enhance 
the localization length $\xi $($\equiv \gamma_2^{-1}$)  of the electronic states
around $|E|< 1$, although the largest Lyapunov exponent $\gamma_1$ 
does not almost change by the effects.

Figure 2(b), 2(c) and 2(d) show 
 the nonnegative Lyapunov exponents in real DNA sequences of 
(b)HCh-22, (c) bacteriophages of E. coli and (d)Histon proteins.
In the case of HCh-22, we used two sequences with $N=10^5$, 
extracted from the original large DNA sequence.
The result shows the Lyapunov exponents do not depend the details of the difference 
of the sequence in  HCh-22.  
Although the weak long-range
correlation has been observed in  HCh-22 as mentioned in introduction, 
it does not affect on the localization property.   The sequences we used are almost symmetric
($R_{GC} \sim 0.5$).
  
\begin{figure}[!h] 
\begin{center} 
\includegraphics[scale=.7]{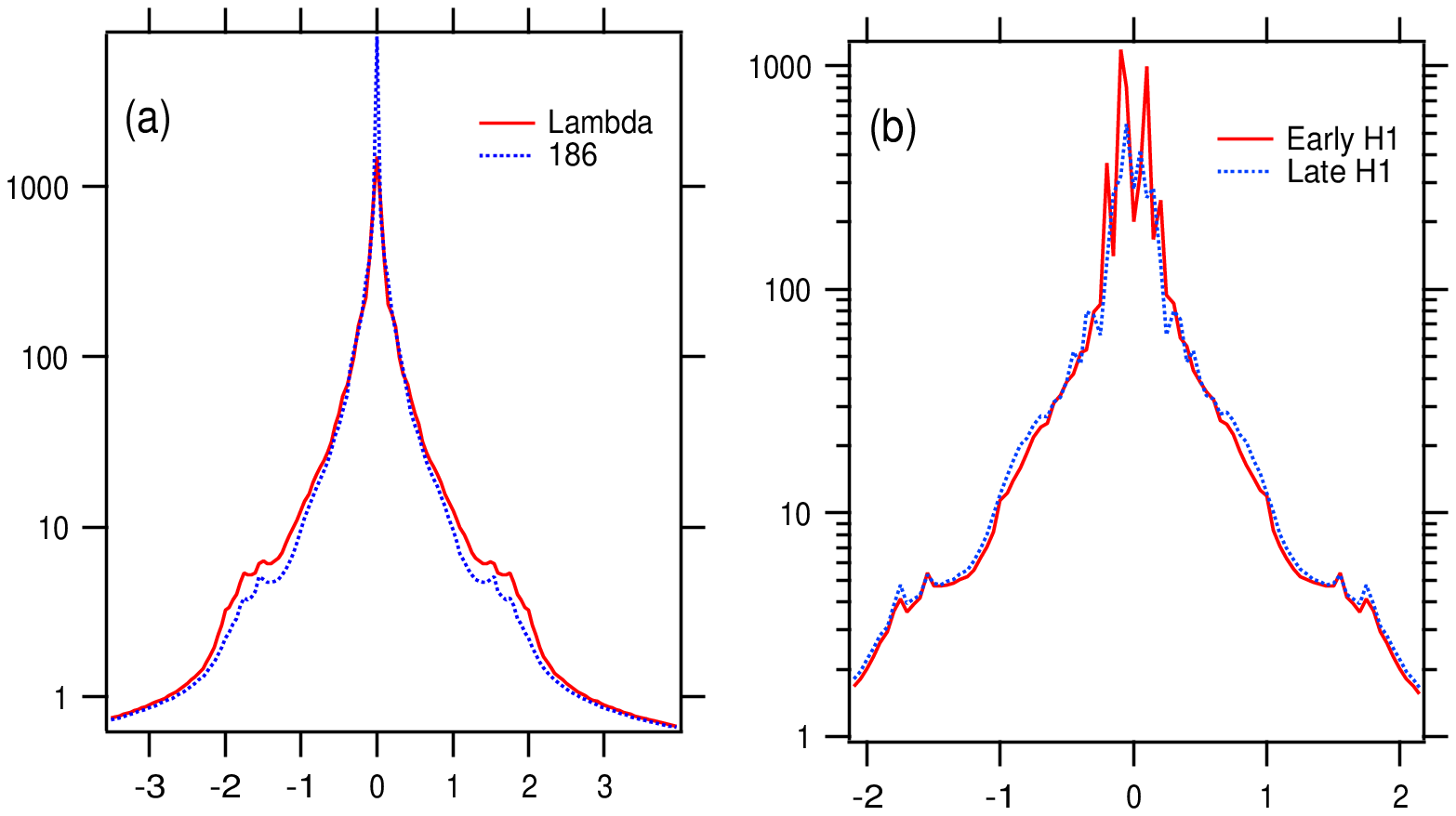}　
 \caption{Localization length $\xi(= \gamma_2)$ as a function of energy in 
the ladder model. (a) bacteriophages (phage-$\lambda$, phage-186), (b)early histone H1 and late histone H1.   
The parameters are same to those in Fig.2.}
\end{center} 
\end{figure}

In Fig.2(c) and (d) the least nonnegative Lyapunov exponents are influenced by the 
difference in the sequence. 
The localization length $\xi_{d}=1/\gamma_d $ defined by the 
least nonnegative Lyapunov exponent for the bacteriophages and histon H1 are shown in Fig.3(a) and (b). 
Apparently the localization length of the phage-$\lambda$ is larger than phage-186. 
Moreover, it seems that 
the difference between sequence of  early histon H1 and late histon H1 effects on 
the resonance structure around $E=|2|$, 
although the difference does not change 
the localization property vicinity of the band center.  



\begin{figure}[!h] 
\begin{center} 
\includegraphics[scale=.8]{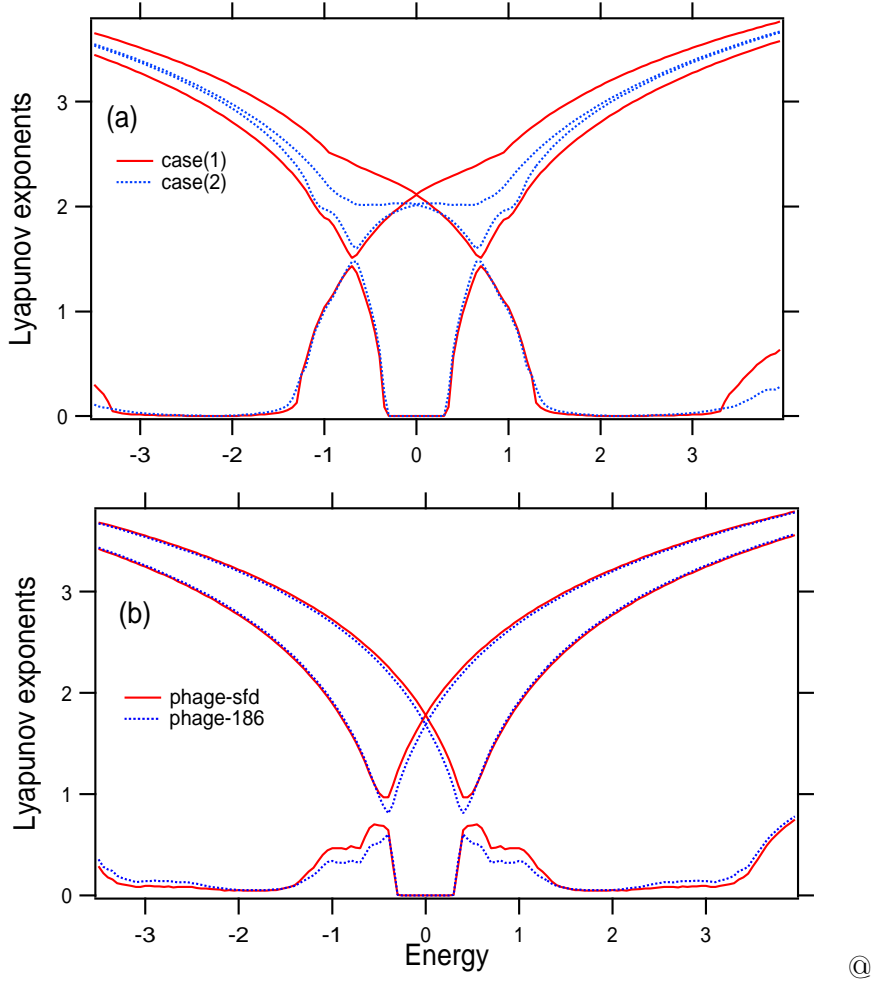}　
 \caption{Lyapunov exponents ($\gamma_1, \gamma_2, \gamma_3$) as a function of energy in 
the three-chain model. (a) modeified Bernoulli model, (b)bacteriophages of E.coli (phage-fd, phage-186). 
The parameters are same as ones in Fig.2 except for on-site energies of C-chain ($C_{n,n}=0$).}
\end{center} 
\end{figure}

Furthermore, we have confirmed that almost similar property to the double-chain model 
have been observed in the three-chain model.
 Figure 4 shows the energy dependence of the Lyapunov exponents ($\gamma_1,\gamma_2,\gamma_3$) 
in the three-chain model. In Fig.4(a) the result for asymmetric modified Bernoulli system is shown.
Appearently, the correlation and/or asymmetry of the sequence effect a change in 
the second and third Lyapunov exponent. In contrast, although the global feature of $\gamma_1$ is almost
unchanged,  the local structure of the energy dependence is changed by the change of $B_0$.
Figure 4(b) shows the results in phage-fd and phage-186 in the three-chain model.
It is found that the structure of the energy dependence around $|E|<2$ is different from that in
the artificial sequence by the modified Bernoulli map.

\section{Summary and Discussion}
We introduced two-chain and three-chain models  
as  simple models for electronic property in the double strand of DNA.
We numerically investigated the correlation effect on 
the localization property of the one-electronic states 
in the disordered two-chain (ladder) and three-chain models  
with the long-range structural correlation by means of asymmetric modified Bernoulli model
and some real DNA sequences.
As a result, the correlation enhances the localization length ($\gamma_2^{-1}$)  
around $|E|<1$, although the $\gamma_1$ does not almost change.
In addition to the correlation effect, the asymmetry 
of the sequence also enhances the localization length. 
Almost similar property to the two-chain model 
have been observed in the three-chain model.

The relation between the correlation length of the DNA sequence and the evolutionary 
process is suggested \cite{isohata03,ohno70}.
Moreover, it is interesting if the localization property would be related to the evolutionary 
process.
 Up to now, although  we consider the effect of the two points correlation of the sequence on the 
localization,  the relation between the general complexity of the sequence and the localization property
is also very interesting future problem \cite{badi97,yamada05}.

Finally, it should be noted that although in the present paper we have focused on 
the localization  properties of the electronic states at zero temparature, 
with enhasis on the ladder geometry by the backbone and correlation of the DNA sequence, 
for the sake of simplicity, 
in the experiment of the conductance property of the DNA,  both the temperature effect and 
the temperature dependence become important.
Indeed, the finite temperature can also reduce the effective system size and leads to 
the changes in the transport property.

The author would like to thank Dr. Kazumoto Iguchi for 
stimulating and useful discussions.


\end{document}